# Ultrafast Spin-To-Charge Conversion at the Surface of Topological Insulator Thin Films

*Xinbo Wang, Liang Cheng, Dapeng Zhu, Yang Wu, Mengji Chen, Yi Wang, Daming Zhao, Chris B. Boothroyd, Yeng Ming Lam, Jian-Xin Zhu, Marco Battiato, Justin C. W. Song, Hyunsoo Yang\*, and Elbert E. M. Chia\**

Dr. X. Wang, Dr. L. Cheng, Dr. D. Zhao, Prof. M. Battiato, Prof. J. Song, Prof. E.E.M. Chia
Division of Physics and Applied Physics, School of Physical and Mathematical Sciences, Nanyang Technological University, 21 Nanyang Link, 637371, Singapore
E-mail: elbertchia@ntu.edu.sg

Dr. D. Zhu, Dr. Y. Wu, M. Chen, Dr. Y. Wang, Prof. H. Yang
Department of Electrical and Computer Engineering, National University of Singapore, 117576, Singapore
E-mail: eleyang@nus.edu.sg

Dr. C. B. Boothroyd, Prof. Y.M. Lam
School of Materials Science and Engineering, Nanyang Technological University, 50 Nanyang Avenue, 639798, Singapore

Dr. J. Zhu
Theoretical Division and Center for Integrated Nanotechnologies, Los Alamos National Laboratory, New Mexico 87545, USA

Dr. X. Wang
Beijing National Laboratory for Condensed Matter Physics, Institute of Physics, Chinese Academy of Sciences, Beijing 100190, China

Prof. M. Battiato
Institute of Solid State Physics, Vienna University of Technology, 1040 Vienna, Austria

Prof. J. Song
Institute of High Performance Computing, Agency for Science, Technology, and Research, Singapore 138632

Topological insulators (TIs) provide a fascinating example where strong spin-orbit interaction leads to the formation of spin-momentum-locked (topological) surface states,[1,2] and are anticipated to mediate superior spin-charge conversion (SCC).[3,4] As a result, employing conventional steady-state transport set-ups, TIs have been exploited to exert spin orbit torque on the adjacent ferromagnetic (FM) layers,[5-12] as well as generate charge current from spin current.[13-19] However, these transport studies involved contact fabrication, required complicated data analysis to extract the spin-related component, and their signal is in the DC regime. Direct access to the *dynamics* of SCC – a crucial ingredient in optimizing TI based spintronic devices – remains conspicuously missing. The difficulty of tracking SCC dynamics stems from the need to address both the ultrafast timescales involved as well as obtain spin resolution simultaneously.

In this work, we probe the dynamical processes of SCC in $Bi_2Se_3$/Co heterostructures directly for the first time, via ultrafast spin injection and terahertz (THz) emission spectroscopy. Ultrafast spin injection from ferromagnetic Co enables transient spin currents to be injected.[20-23] At the same time, given the challenging timescales involved, THz emission spectroscopy – the central signature we track — is a powerful optical tool, which offers a contact-free amperemeter with sub-picosecond resolution. As we describe below, this technique combined with a systematic exhaustion of the experimental parameters of the $Bi_2Se_3$/Co heterostructures and our ultrafast setup (thickness, temperature, Co magnetization direction, optical pulse polarization, etc.), enables an unprecedented detailed window into the key parameters of SCC including the timescale as well as the spatial locality of ultrafast spin-to-charge conversion.

In particular, we find a highly efficient THz emission (mediated by ultrafast spin injection) from the $Bi_2Se_3$/Co heterostructures. Strikingly, this far outstrips (by up to an order of magnitude) the emission of either bare $Bi_2Se_3$ or Co films taken individually, indicating the highly efficient SCC in the $Bi_2Se_3$/Co stack, as well as the central role SCC plays in the optical response of the heterostructure. Crucially, the THz signal in the $Bi_2Se_3$/Co stack exhibits a significant enhancement when the thickness of $Bi_2Se_3$ increases from four to six quintuple layers. Coinciding with the appearance of well-defined surface states,[24] this indicates the dominant role of surface states in the SCC processes.

In addition, varying pump polarization enables a direct tracking of the SCC process. We identify the fundamental timescale for ultrafast SCC formation in $Bi_2Se_3$/Co to be ~0.12 ps.

This sets a characteristic cut-off time for SCC processes in TI spintronic heterostructures. Our results provide fundamental insights into the ultrafast SCC processes and indicate the central role of topological surface states in the very efficient SCC at room temperature for THz opto-spintronic applications.

**Dynamical response of pure $Bi_2Se_3$ thin films.** In order to methodically delineate the THz emission signals in our $Bi_2Se_3$/Co heterostructures (see below), we first describe the optical response (focusing on its THz response) of pure $Bi_2Se_3$ thin films grown by molecular beam epitaxy (MBE). As we will discuss, this enables us to clearly isolate the new dynamics that arises due to the TI/FM heterostructure stack from the response of either the pure TI or pure FM by themselves.

$Bi_2Se_3$ thin films samples were setup in a transmission geometry, as shown in **Figure 1**a, where an 800-nm pump beam with 60 mW power impinged on the samples at normal incidence, and only the vertical components of the emitted THz pulses were detected by electro-optic sampling. The typical THz electric field waveforms emitted from $Bi_2Se_3$ films are plotted in Figure 1b. The spectra of the detected THz signals, as shown in the inset of Figure 1b, covers the frequency range up to 3 THz, which is limited by the electro-optic detection of 1-mm-thick ZnTe crystal. A $SiO_2$ capping layer is essential to protect the $Bi_2Se_3$ film from the atmosphere, and prevent the evaporation of the thin film under pulsed-laser illumination due to the weak van der Waals type coupling between the two neighboring quintuple layers (QLs, 1 QL~1 nm).[25] The detected THz traces show a distinct dependence on the pump polarization. Figure 1c shows the THz peak amplitude as a function of pump polarization angle, which can be fitted well by a sinusoidal function with a period of 180°. In addition, the dependence of THz amplitude on sample azimuthal angle reveals a characteristic three-fold symmetry (see Figure S2), which is associated with the crystalline symmetry of $Bi_2Se_3$.[26] These are in good agreement with the expectations from the shift current mechanism (see Supporting Information Section S2, for further exclusion of other mechanisms).[27,28] Additionally, we note that the dependence of the THz peak amplitude on the incident pump power, as indicated in the inset of Figure 1d, shows a nonlinear behavior, which we attribute to saturable absorption (Supporting Information Section S3).[29]

In principle, the shift current is a second-order nonlinear effect, which requires the breaking of spatial inversion symmetry,[30-32] and hence can only be excited on the surface of three-dimensional TIs.[33-35] Consequently, the amplitude should be independent of the thickness of the $Bi_2Se_3$ film. Indeed, the THz peak amplitude is roughly constant for the samples above 6 QL, as illustrated in Figure 1d. Interestingly, for the 4-QL sample, the measured THz signal is about one order of magnitude smaller than that from the films above 6 QL. It has been reported that the shift current in $Bi_2Se_3$ originates from the transient charge transfer along the Se-Bi bonds involving the surface-states-related optical transitions.[27] However, for the ultrathin $Bi_2Se_3$ film with a thickness below 6 QL, the coupling between the top and bottom surface opens a gap in the surface-state dispersion,[24] and hence suppresses the surface-states-related optical transitions.[36] Therefore, the shift current and resulting THz emission decrease considerably. This observation indicates that the THz emission from pure $Bi_2Se_3$ film is a surface-dominated response.

**THz emission from $Bi_2Se_3$/Co heterostructures.** We now turn to investigate the THz response of $Bi_2Se_3$/Co heterostructures. As we show below, the dynamics of this TI/FM heterostructure starkly contrasts with that of the TI by itself. The experiments are performed under identical conditions as that for the pure $Bi_2Se_3$ films, except that an in-plane external static magnetic field ~800 Oe is applied to orient the Co magnetization, as sketched in **Figure 2**a. The THz waveform emitted from $Bi_2Se_3$/Co heterostructure, together with that from pure $Bi_2Se_3$ and Co films, is shown in Figure 2b.

Strikingly, the THz emission from the heterostructure is significantly enhanced after the deposition of a Co layer. Indeed, for the same TI film thickness, the THz emission amplitude from the $Bi_2Se_3$/Co heterostructure (orange curve) is about an order of magnitude larger than that of a pure Bi$_2$Se$_3$ film by itself (blue curve). We note that the detected THz emission from a 3-nm Co film by itself (green curve) is weak; this indicates that the laser-induced demagnetization in the magnetic film contributes negligibly to THz emission in TI/FM heterostructures.[37,38] Furthermore, in contrast to that of the pure Bi$_2$Se$_3$ film, the THz emission amplitude from the heterostructure obeys a linear power dependence up to high power (> 40 mW), as illustrated in the inset of Figure 2d. These observations indicate that THz emission from the heterostructure cannot be explained by dynamics from that of the pure TI or FM by themselves.

Instead, as we now explain, it arises from the novel coupled spin-charge dynamics in the Bi$_2$Se$_3$/Co heterostructure. In order to understand this, we note that under the illumination of 800 nm laser beam, a spin-polarized current is set up at the interface of the Bi$_2$Se$_3$/Co heterostructure due to the different mobilities of the majority- and minority-spin *sp*-band electrons in the Co layer.[20] Due to the existence of strong SOC in Bi$_2$Se$_3$, the ultrafast injected spin current can be converted to an ultrafast charge current via either the inverse Edelstein effect (IEE) on the surface or the inverse spin Hall effect (ISHE) in the bulk.[39-42] This SCC conversion process can lead to the large-amplitude THz waveforms that we detect from the TI/FM heterostructure. We note that the SCC transient charge currents generated by both IEE and ISHE flow in the same direction — since $\boldsymbol{j_c} \propto \hat{\boldsymbol{z}} \times \boldsymbol{\sigma}$ in IEE[13,41,43] and $\boldsymbol{j_c} \propto \boldsymbol{j_s} \times \boldsymbol{\sigma}$ in ISHE,[39,42] where $\boldsymbol{j_c}$ is the charge current, $\boldsymbol{j_s}$ is the spin current, $\boldsymbol{\sigma}$ is the direction of the spin polarization, and $\hat{\boldsymbol{z}}$ is the unit vector normal to the TI/FM interface (which is parallel to $\boldsymbol{j_s}$). As a result, the polarization of the emitted THz pulses from both ISHE and IEE also point in the same direction.

To confirm the central role of SCC in the Bi$_2$Se$_3$/Co stack THz emission, we oriented the magnetization in the Co layer using an in-plane magnetic field — this enables the tuning of the spin-polarization injected into the TI film. As shown in Figure 2c, the polarity of the emitted

THz pulses reverses when either (1) the direction of magnetic field was reversed, or when (2) the pump beam was incident on the opposite side of the heterostructure, as expected from SCC (IEE and/or ISHE) THz emission mechanism. While (1) displays the spin origin of the $Bi_2Se_3$/Co stack THz emission, (2) displays its spin-current dependence, since the spin current flips its direction relative to the lab frame when the sample is pumped from the back.

Indeed, since the spin polarization $\sigma$ is determined by the external magnetic field, the abovementioned two SCC mechanisms also predict that the measured ($x$-projected) THz electric field, obeys $E_{THz}^x \propto |\boldsymbol{j_s} \times \boldsymbol{\sigma}|_x \propto sin\phi$, where $\phi$ is the magnetic field angle with respect to the positive $x$-axis. As shown in Figure 2d, the THz peak amplitude is locked to the magnetic field direction and exhibits a one-fold sinusoidal behavior, and allows direct magnetic field control of THz emission. This, therefore, demonstrates a novel spin-charge dynamics that is enabled by SCC in TI/FM heterostructures.

**Separation of THz waveforms and SCC timescale.** Pump-polarization-dependent measurements provide a further means of quantifying the contributions to the THz emission from the $Bi_2Se_3$/Co heterostructures. In contrast to that from pure $Bi_2Se_3$ films, we observe only small changes in the amplitude and phase of the measured THz waveform as a function of pump polarization angle, as shown in **Figure 3**a. This is fully consistent with a dominant THz emission contribution from SCC that is independent of pump polarization.[21-23] Figure 3b plots the THz electric field at $t = 0$ ps as a function of the pump polarization angle, showing a small-amplitude oscillation superposed on top of a large dc offset. As explained in the last section, the small-amplitude polarization-dependent contribution can be ascribed to the THz generation from the $Bi_2Se_3$ film itself.

On the basis of these characteristics, we can separate out the THz waveform contributed by (a) the TI film alone (that is via the shift-current mechanism), $E_{TI}(t)$, and (b) from SCC in the TI/FM heterostructure, $E_{SCC}(t)$. This is done by performing for (a) $E_{TI}(t) = [E(t, \theta = 0°) - E(t, \theta = 90°)]/2$, and for (b) $E_{SCC}(t) = [E(t, \theta = 0°) + E(t, \theta = 90°)]/2$. The resulting delineated THz waveforms are shown in Figure 3c. For comparison, the THz waveform emitted from pure 10-QL $Bi_2Se_3$ film is also plotted. Notice that in the $Co/Bi_2Se_3$ heterostructure, the THz signal contributed by the shift-current mechanism is smaller than that from pure $Bi_2Se_3$ — this can be attributed to the reflection or absorption of the 800-nm pump beam by the Co film. The extracted TI component shows a very similar temperature dependence to that of pure $Bi_2Se_3$ film, as shown in Figure S7, giving further evidence that they share the same origin (Supporting Information Section S7).

Interestingly, this delineation enables a temporal difference between these two extracted THz waveforms to be discerned — the SCC-component lags the TI-component (shift current) by ~0.12 ps. This can be seen more clearly in the plot of normalized THz waveforms in the inset of Figure 3c.

The significance of this delay can be appreciated by first noting that to a good approximation, we can assume that the 800-nm light hits the surface of the Co and $Bi_2Se_3$ layer simultaneously due to the ultrathin Co layer. Upon optical excitation, the shift current originates from the instantaneous shift of electron density along the Se-Bi bond,[27] while it takes a longer time to generate a spin current in the FM layer, diffuse into the $Bi_2Se_3$ film, and then be converted to a charge current.[22,23] To quantify the spin transport in the FM layer, we vary the Co thickness from 2 to 16 nm while keeping the thickness of $Bi_2Se_3$ fixed at 10 QL. The amplitude of the THz signal contributed by the SCC component is maximum at ~3 nm, and then decreases with increasing Co thickness (see Figure S3), which we attribute to the smaller amounts of 800-nm light reaching the $Co/Bi_2Se_3$ interface due to the small penetration depth in Co (13 nm)[44] and hence smaller injection of the laser-driven spin current.[22,23] Figure 3d shows that the extracted temporal delay is independent of the thickness of Co layer. This is consistent with the picture of superdiffusive spin transport in Co, where the timescale of the multiplication of carriers is only dependent on the material (Supporting Information Section S4). Therefore, the temporal difference of ~0.12 ps corresponds to the time needed for the formation of a maximum charge current from the peak of the laser pulse, and thus characterizes the SCC processes in the Bi$_2$Se$_3$/Co heterostructure. This value is of the same order as that deduced by theoretical simulations of the spin dynamics (0.1 – 0.3 ps) in heavy-metal/FM heterostructures.[22,23]

**Thickness dependence of THz emission from Bi$_2$Se$_3$/Co heterostructures.** In order to verify whether the observed SCC is dominated by IEE on the surface or ISHE in the bulk, we perform measurements on different thicknesses of Bi$_2$Se$_3$ film while fixing the Co thickness at 4 nm. **Figure 4**a shows the THz peak amplitude of the extracted SCC component as a function of Bi$_2$Se$_3$ thickness. At 4 QL, the emitted THz signal is relatively small, while it increases by about one order of magnitude as we go from 4 to 6 QL. We see the same thickness dependence of THz emission in pure Bi$_2$Se$_3$ films with thicknesses from 4 to 10 QL (shown in Figure 1d), indicating therefore that the THz emission mechanism in TI/FM is also a surface-dominated response. Although the Bi$_2$Se$_3$ film may be doped by the metallic Co layer (Supporting Information Section S1), the dramatic enhancement of the THz signal at 6 QL can be still ascribed to the formation of gapless surface states,[24] as has been discussed earlier in pure Bi$_2$Se$_3$ films.

On the other hand, if ISHE in the bulk were to play a major role in THz emission of TI/FM heterostructures, the detected THz signal can be written as $E_{THz}^x \propto \lambda_{sf} tanh(d/2\lambda_{sf})$, where $\lambda_{sf}$ is the spin diffusion length and $d$ is the thickness of Bi$_2$Se$_3$ film.[18,40,45] However, this spin diffusion model cannot account for the Bi$_2$Se$_3$ thickness dependence of THz peak amplitude (Supporting information Section S5). Therefore, we conclude that the SCC processes in TI/FM heterostructures are dominated by the topological surface states, though we cannot completely exclude a possible contribution from ISHE in the bulk. For the thicker samples (above 10 QL), the extracted THz signal decreases with increasing thickness. This probably originates from the carrier absorption in the bulk since the THz transients are generated at the interface of the heterostructure. Figure 4a also shows the electric-field transmittance of pure Bi$_2$Se$_3$ films at 1

THz (corresponding to the peak of THz waveforms) upon optical excitation, which decreases with increasing film thickness (Supporting Information Section S6). The observed agreement between the thickness dependence of the THz signal and the pump transmittance (for samples above 4 QL) suggests that the suppression of the THz amplitude is attributed to the absorption of THz radiation by both the intrinsic bulk carriers and the photo-generated free carriers in $Bi_2Se_3$ films.

**Temperature dependence.** We also investigated the temperature dependence of THz emission from both the $Bi_2Se_3$ (10 QL) film and the $Bi_2Se_3$/Co (4 nm) heterostructures with different $Bi_2Se_3$ thicknesses. Cooling the pure $Bi_2Se_3$ film from room temperature down to 10 K increases the emitted THz signal, which we attribute to a higher carrier mobility at low temperature (Supporting Information Section S7). In contrast, as shown in Figure 4b, the THz peak amplitude of the extracted SCC component does not change with temperature for four representative samples with different $Bi_2Se_3$ thicknesses. In this temperature range (10 – 300 K), the spin diffusion length of the Co layer (38 – 59 nm) is much larger than the thickness of Co layer used in our case,[46] implying a constant spin current injection from the Co layer into the TI at different temperatures. Therefore, the temperature-insensitive THz emission implies that the SCC efficiency via IEE is also independent of temperature. This is consistent with previous work on the temperature evolution of IEE in metallic Ag/Bi junctions.[47] In TIs, IEE originates from the spin-momentum-locked surface states, which is a result of strong SOC.[1,2] As a relativistic effect, it is expected to be temperature-independent.[48] Consequently, the resulting THz signal contributed by the IEE should be temperature insensitive, in good agreement with our experimental results, confirming once again the dominant contribution from the topological surface states to the observed SCC in TI/FM heterostructures.

The spin-momentum-locked topological surface states have been predicted to possess superior SCC efficiency.[1,3,4] However, previous reported values of the SCC efficiency exhibit a large range over several orders of magnitude (from $10^{-4}$ to 0.8).[13-17] In order to evaluate the SCC efficiency in Bi$_2$Se$_3$ (10 QL)/Co (3 nm) heterostructures, we compare its THz emission amplitude to that from the Pt (2 nm)/Co (3 nm) bilayer structure — we choose a Pt thickness of 2 nm to match a single TI surface state thickness of $t \sim 2$ nm,[49,50] since SCC via IEE is restricted to the top surface of Bi$_2$Se$_3$ in the heterostructure. As shown in Figure S8, the THz peak amplitude of Bi$_2$Se$_3$ (10 QL)/Co (3 nm) is ~1.7 times of that from Pt (2 nm)/Co (3 nm) under identical experimental conditions. Since the spin currents are generated by optically exciting an out-of-equilibrium electron distribution in the Co layer (Supporting Information Section S4), it is reasonable to assume that they have similar magnitudes in both heterostructures.[20,22] Consequently, the effective SCC efficiency $\theta_{eff} = j_c/j_s$ is a function only of charge current and hence THz signal amplitude. As a rough estimate, we use a measured spin Hall angle of ~0.03 for the 2-nm-thick Pt film.[51-54] Thus, we obtain $\theta_{eff} \sim 0.03 \times 1.7 \sim 0.05$ for 10-QL Bi$_2$Se$_3$ film. Using $\lambda_{IEE} = \frac{1}{2}\theta_{eff} \cdot t$ as defined in Equation S7 and $t \sim 2$ nm, we obtain the IEE length $\lambda_{IEE} \sim 50$ pm charactering the SCC efficiency of the surface states, which is in good agreement with the spin-pumping measurements in TI/ferromagnetic-insulator heterostructures (35 – 76 pm).[18,19] Note that if we use a larger measured spin Hall angle for the 2-nm-thick Pt film,[51-54] as well as include the interfacial spin transparency,[54-57] we will obtain a larger SCC efficiency for Bi$_2$Se$_3$ (see Supporting Information Section S8 for further discussion on these points).

Remarkably, the amplitude of the THz emission from the Bi$_2$Se$_3$ (10 QL)/Co (3 nm) heterostructure is only slightly smaller than that from the Pt (6 nm)/Co (3 nm) bilayer structure (see Figure S8), and the latter has been demonstrated to be a very efficient broadband THz emitter comparable to the standard ZnTe THz emitter.[21,23,58] In this regard, the temperature-insensitive SCC efficiency suggests promising applications of topological surface states in THz spintronic devices at room temperature.

In conclusion, we have systematically demonstrated, for the first time, highly-efficient ultrafast spin-injection and spin-to-charge conversion in topological-insulator/ferromagnet heterostructures using THz emission spectroscopy. Locked to the Co magnetization direction, THz emission from $Bi_2Se_3$/Co heterostructures far outstrips the emission of either $Bi_2Se_3$ or Co individually. This underscores the central role the novel TI spin-charge dynamics play in response of the heterostructure. While we have focused on the THz response of $Bi_2Se_3$/Co to elucidate its SCC dynamics, large non-equilibrium spin injection into a TI afforded by our methodology unlocks easy access to a range of spin-type responses previously difficult to realize in TIs, including tunable generation of ultrafast spin and charge current in the bulk and/or surface. Perhaps most tantalizing is the fast timescale in which efficient SCC in $Bi_2Se_3$ occurs within ~0.12 ps. This paves the way for high speed and ultra-thin opto-spintronic devices with operating speeds up to several terahertz.

**Experimental Section**

*Sample fabrication*: High-quality thin films of the topological insulator $Bi_2Se_3$ with thicknesses of 4, 6, 8, 10, 20, and 40 QLs were grown by a two-step deposition process using MBE on c-axis $Al_2O_3$ substrate (0.5 mm). The samples were transferred through air into the sputtering chamber for capping layer and magnetic layer deposition. For pure $Bi_2Se_3$ films, the samples were protected by a $SiO_2$ capping layer (4 nm). For the $Bi_2Se_3$/Co heterostructures, a nanometer-thick Co layer (2 to 16 nm) was deposited onto the $Bi_2Se_3$ films by sputtering and then capped with the same $SiO_2$ capping layer to prevent oxidation of the magnetic layer. For comparison, a pure Co layer (3 nm) on an $Al_2O_3$ substrate with the same capping layer was also prepared. Please see Supporting Information Section 1 for more sample information.

*THz emission setup*: A Ti:sapphire regenerative amplifier (Coherence Legend; center wavelength 800 nm; duration 50 fs; repetition rate 1 kHz) was employed in this work. The beam was split into two parts: one for the pump beam and the other one for THz detection, by an optical beam splitter. The pump beam with an incident power of 60 mW and a spot size 7 mm (0.156 mJ/cm$^2$) impinged on the samples at normal incidence from the sample side. Any residual pump beam after the sample was blocked by a piece of 1/16-inch-thick high-density polyethylene (HDPE), which is transparent in the THz frequency range. The THz pulses generated were projected onto the x-axis (vertical) by using a wire-grid THz polarizer, and then collected and focused by two parabolic mirrors onto a 1-mm-thick <110> ZnTe crystal for free-space electro-optic sampling. A cryostat was used in the temperature-dependent experiments, in which the magnets were fixed along the y (horizontal) direction. The remaining experiments were performed at room temperature in a dry air atmosphere. A half-wave plate and a pair of permanent magnets in the dipole configuration were mounted on rotation stages to vary the polarization of pump pulses and to align the magnetization of the Co layer, respectively. A visible wire grid polarizer for the 800-nm beam was placed before the half-wave plate. The magnetic field at the sample position was ~800 Oe.

**Supporting Information**

Supporting Information is available online.

**Acknowledgements**

X.W., L.C. and D.Z. contributed equally to this work. We acknowledge funding from the A*Star PHAROS Programme on Topological Insulators (SERC Grant No. 152 74 00026) and 2D Materials (SERC Grant No. 152 70 00016), and the Singapore Ministry of Education AcRF Tier 2 (MOE2015-T2-2-065). The electron microscopy work was performed at the Facility for Analysis, Characterization, Testing and Simulation (FACTS), Nanyang Technological University, Singapore. J.-X.Z acknowledges the support of the U.S. DOE BES Coe Program E3B7. J.C.W.S. acknowledges the support of the Singapore National Research Foundation (NRF) under NRF fellowship award NRF-NRFF2016-05 and the Nanyang Technological University through a SUG Grant. M.B. acknowledge the financial support of the Austrian Science Fund (FWF) through a Lise Meitner grant M1925-N28 and by a Nanyang Technological University NAP-SUG Grant. The work was supported in part by the Center for Integrated Nanotechnologies, a U.S. DOE BES user facility.

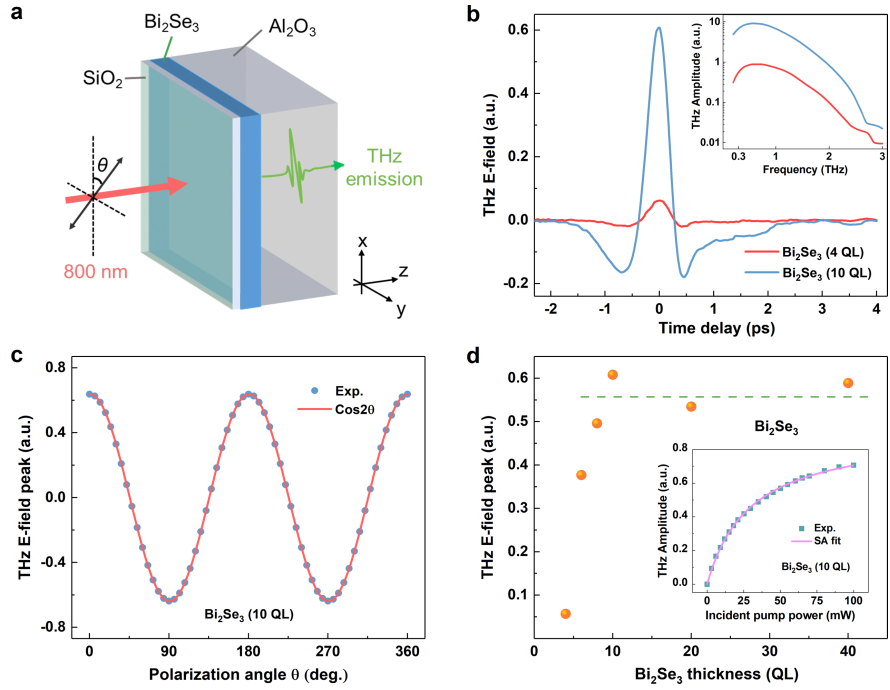

**Figure 1. THz emission from Bi$_2$Se$_3$ thin films.** a) Schematic illustration of THz emission from Bi$_2$Se$_3$ films. The sample is excited by femtosecond laser pulses (800 nm, 60 mW) from the sample side, and the vertical component of the emitted THz pulses is detected by electro-optic sampling. θ is polarization angle of pump beam with respect to the $x$-axis. b) Typical THz waveforms emitted from Bi$_2$Se$_3$ films with different thicknesses. The inset shows the corresponding spectra. c) Peak amplitude of THz waveforms from 10-QL Bi$_2$Se$_3$ film as a function of polarization angle θ. The solid line shows a sinusoidal fit. d) Peak amplitude of the THz signal as a function of Bi$_2$Se$_3$ thickness. The dashed line is a guide to the eye. The inset shows the power dependence of the THz emission from the 10-QL Bi$_2$Se$_3$ film, exhibiting saturation behavior. The solid line shows the fit with the saturable absorption model.

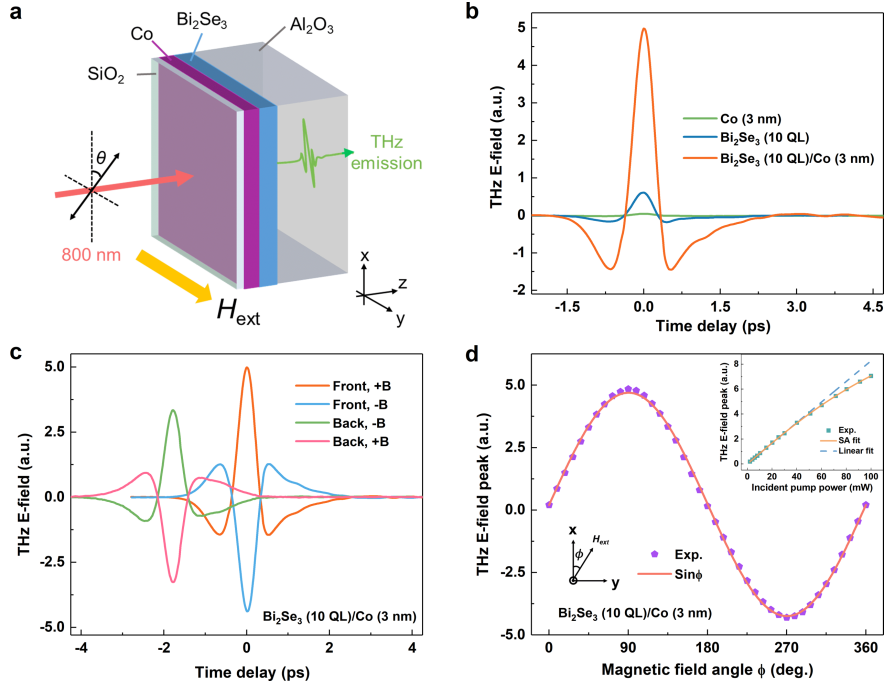

**Figure 2. THz emission from Bi$_2$Se$_3$ (10 QL)/Co (3 nm) heterostructure.** a) Sketch of the experimental geometry for the Bi$_2$Se$_3$/Co heterostructures. An external magnetic field ***H**$_{ext}$* is applied to align the Co magnetization. b) THz waveforms generated from Co (3 nm) film, Bi$_2$Se$_3$ (10 QL) film, and the Bi$_2$Se$_3$ (10 QL)/Co (3 nm) heterostructure. c) THz waveforms emitted from the Bi$_2$Se$_3$ (10 QL)/Co (3 nm) heterostructure measured with front and back sample excitation and reversed magnetic field. Note that the differences in amplitude and time delay of the THz waveforms on pumping the samples from opposite sides originate from the different refractive indices of the sapphire substrate in the 800-nm and THz frequency range. d) The peak amplitude of the THz pulses emitted from the Bi$_2$Se$_3$ (10 QL)/Co (3 nm) heterostructure as a function of magnetic field angle ϕ. The solid line shows a sinusoidal fit. Inset: Power dependence of the THz emission from the Bi$_2$Se$_3$ (10 QL)/Co (3 nm) heterostructure. The solid line shows a fit with a saturable absorption model. The dashed line shows a linear fit to the data below 40 mW.

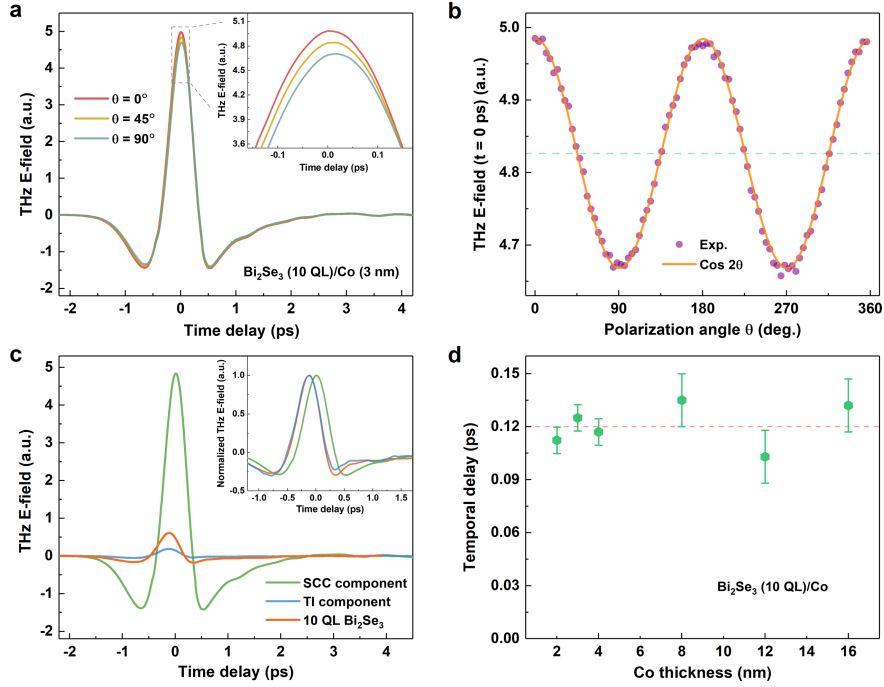

**Figure 3. Separation of THz waveforms and SCC timescale.** a) THz waveforms generated from the Bi$_2$Se$_3$/Co heterostructure under a range of pump polarization angles. Inset: enlargement of the THz waveforms near the peak. b) The electric field of the THz traces at $t = 0$ ps as a function of pump polarization angle θ. The solid line shows a sinusoidal fit with a large offset as indicated by the dashed line. c) The extracted THz waveforms contributed by the SCC and TI components. The THz signal from pure Bi$_2$Se$_3$ (10 QL) film is also shown for comparison. The inset shows the normalized waveforms. d) Extracted temporal delay between the SCC and TI components as a function of Co thickness. The dashed line is a guide to the eye.

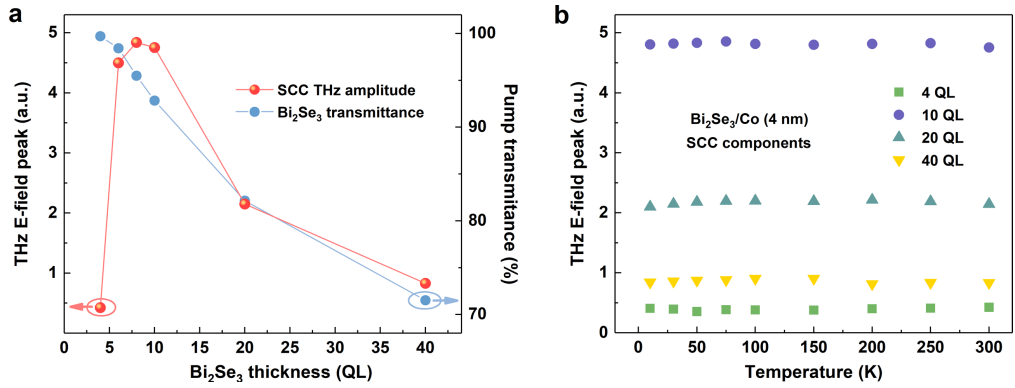

**Figure 4. Bi$_2$Se$_3$ thickness and temperature dependence of THz emission**. a) THz peak amplitude of the extracted SCC component from Bi$_2$Se$_3$/Co (4 nm) heterostructures (red) and THz electric field transmittance amplitude of Bi$_2$Se$_3$ films under a pump power of 30 mW (blue), as a function of Bi$_2$Se$_3$ thickness. b) Temperature dependence of the THz peak amplitude of the extracted SCC component from Bi$_2$Se$_3$/Co (4 nm) heterostructures with different Bi$_2$Se$_3$ thicknesses.